\pdfoutput=1
\documentclass[runningheads]{llncs}
\usepackage{graphicx}
\usepackage{epsfig}
\usepackage{amsmath}
\usepackage{amssymb}
\usepackage{mathrsfs}
\usepackage{tabu}
\usepackage{booktabs}
\usepackage{graphbox}
\usepackage{float}
\usepackage{soul}
\usepackage{xcolor}
\usepackage{color, soul}
\usepackage[belowskip=-15pt,aboveskip=0pt]{caption}
\usepackage{tabularx}

\usepackage{textcomp}    

\usepackage{array}
\newcolumntype{L}{>{\centering\arraybackslash}m{3cm}}

\begin{document}
	\title{Multimodal Medical Volume Colorization from 2D Style}
	%
	%
	\author{Aradhya Neeraj Mathur\inst{1} \and
		Apoorv Khattar\inst{1} \and
		Ojaswa Sharma\inst{1}\orcidID{0000-0002-9902-1367}}
	\authorrunning{Mathur et al.}
	%
	\institute{Department of Computer Science and Engineering, IIIT Delhi, India\\
		\email{\{aradhyam,apoorv16016,ojaswa\}@iiitd.ac.in}}
	\maketitle              
	\begin{abstract}
	
    Colorization involves the synthesis of colors on a target image while preserving structural content as well as the semantics of the target image. This is a well-explored problem in 2D with many state-of-the-art solutions. We propose a novel deep learning-based approach for the colorization of 3D medical volumes. Our system is capable of directly mapping the colors of a 2D photograph to a 3D MRI volume in real-time, producing a high-fidelity color volume suitable for photo-realistic visualization. Since this work is first of its kind, we discuss the full pipeline in detail and the challenges that it brings for 3D medical data. The colorization of medical MRI volume also entails modality conversion that highlights the robustness of our approach in handling multi-modal data.
		
	\end{abstract}
	\section{Introduction}
	Image colorization enables transferring radiometric characteristics of one image on to another. From an artistic perspective, it has opened new horizons in content creation and has seen rapid strides with the advent of digital media. It has been extensively explored in the 2D domain, however, has not been very successfully applied in 3D. Our current research aims at finding a robust approach to apply photorealistic style transfer to 3D domain and using it for stylizing 3D medical data to improve the volume visualization capabilities of current systems.
	
	Currently, 3D CT and MRI datasets are visualized in color by means of a Transfer Function (TF) that maps intensity values to colors. The creation of such a mapping is mostly manual making it cumbersome and time-consuming. Automatic and semi-automatic synthesis of TFs have been explored in the volume visualization community \cite{ljung2016state}. Colorization can be seen as an effective alternative to using a transfer function, however, it is an ill-posed problem since MRI or CT medical data is a different modality compared to a photograph.
	
	In this work, we propose a novel method to perform modality conversion of a 3D MRI to a photographic volume. While doing so, we also address the challenges of handling 3D data in medical deep learning. We show multi-modal volume colorization using only a single 2D style image. Such a colorization can be used to create better 3D visualizations without the need of creating time-consuming transfer functions. Our approach can be used very well in a straightforward manner to colorize other modalities like 3D CT.
	
	\section{Related Work}
    Here we consider multi-modal colorization where the source and the target images belong to different modalities, which are 3D MRI scans and photographic volumes in our case. There are many approaches for colorization of grayscale images and they mostly differ in the additional input required along with a grayscale image for generating results. 
	
	\subsubsection*{Classical approaches to colorization} Several classical methods look into solving color transfer by treating it a texture transfer problem. Welsh et al. \cite{welsh2002transferring} propose a patch matching algorithm that requires a source image from which color information has to be transferred to the target image. Their approach performs a jittered sampling on the source image to get color patches, finds the closest matching patch from the sampled data and transfers colors onto the target image. Levin et al. \cite{levin2004colorization} view colorization as an optimization problem where given some hints in the form of scribbles, colorization is performed with the constraint that similar intensities should have similar colors. Nie et al. \cite{nie2007optimization} develop on the above idea and use quad-tree image decomposition for cost function minimization.
	
	\subsubsection*{CNN based colorization} With the advent of Convolutional Neural Networks (CNNs), visual tasks have seen rapid strides. Iizuka  et al. \cite{iizuka2016let} leverage the estimation capability of CNNs and perform colorization using both local and global information present in an image. Their approach relies on the classification of an image to generate global priors in the form of class labels that are added to the colorization network for better results. Zhang et al. \cite{zhang2016colorful} propose an automatic colorization algorithm where colorization is treated as a multinomial classification problem. The authors employ a CNN to perform colorization in the CIELab color space by predicting only the chroma channel. Zhang et al. \cite{zhang2017real} propose an interactive colorization algorithm that utilizes user-generated hints and demonstrate the generation of different results for the same grayscale image.
	
	\subsubsection*{Style transfer as a generalization} Style transfer forms an integral part of our proposed method to generate hints for colorizing a volume. Prior to the appearance of neural style transfer, a number of approaches on artistic stylization and non-photorealistic rendering were explored. In order to achieve a more generalized style transfer Hertzmann et al. \cite{hertzmann2001image} perform an example-based style transfer. Gatys et al. \cite{gatys2016image} perform style transfer by minimizing reconstruction losses for feature and style with a pre-trained VGG network \cite{vgg}. Their seminal work demonstrated that CNNs are capable of extracting content information from a photograph and style information from an artwork image to produce a stylized image.  Li et al. \cite{li2017universal} perform image reconstruction along with whitening and coloring transforms (WCT). The authors use VGG convolutions as an encoder and train a decoder symmetrical to VGG to produce photorealistic stylizations. Jamrivska et al. \cite{jamrivska2019stylizing} propose a non deep learning based approach for stylizing videos. For an input video, they require some hand-crafted frames as stylization hints. A patch-based optimization uses a mask guide for an object of interest, a color guide for capturing color information and a positional guide to capture structural information.
	
	\subsubsection*{Modality conversion} For a more realistic colorization, we perform modality conversion on the input MRI volume. GANs have been explored for converting the modality of medical volumes, commonly known as medical image synthesis. Nie et al. \cite{nie2017medical} propose a method to convert MRI to CT using GANs. Their method takes an MRI image and generates the corresponding CT image using a fully convolutional network and use a gradient difference loss function. This work is further extended by Nie et al. in \cite{nie2018medical} where they combine this approach with Auto-Context \cite{tu2009auto} and create context aware GAN to generate 7T MRI from 3T MRI and CT from MRI. In a two-step approach, Guibas et al. \cite{guibas2017synthetic} use a DCGAN to generate synthetic segmentation maps and teach a GAN to generate photorealistic images.
	
	\section{Our approach}
\begin{figure}[htp]
		\centering
		\includegraphics[width=0.75\linewidth]{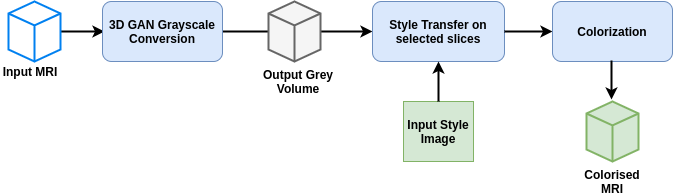}
		\caption{Our colorization pipeline based on modality conversion of input MRI.}
		\label{fig:pipeline_overview}
	\end{figure}

	Our approach enables us to bypass designing a transfer function for color volume visualization. The colorization of MRI volume shows the capability of our approach in handling multi-modal data. In our three-step approach, we use a Generative Adversarial Networks (GANs) \cite{goodfellow2014generative} to convert the modality of the input MRI volume to a corresponding gray cryosection (cryo) volume. This is followed by style transfer on selected slices in the generated gray volume, and a subsequent colorization of the entire volume. The steps of the colorization are summarized in Fig. \ref{fig:pipeline_overview}.
	
	\subsection{Dataset}
	We use Visible Human 2 \cite{ackerman1998visible} dataset which contains unregistered 16-bit MRI and 24-bit color cryo imaging data of human head. It is necessary to perform a registration between the two modalities in order to get correspondence between the two. We perform a rigid registration of MRI with cryo volume using fiducial markers. 
	
	We also perform a marker-based segmentation of the MRI volume into foreground and background. The binary background \emph{segmentation mask} is used in removing background for visualization purposes.
	
	\subsection{MRI to grayscale conversion}
	A direct colorization from MRI to color will not yield good results given substantial differences in the radiometry of MRI imaging and photographs. An MRI response of tissue is different than it is for visible light. That is to say that tissues in an MRI will look different than they will in a photograph. Therefore, the first step of colorization is to convert a given MRI to a corresponding grayscale photographic volume (hereafter referred to as a \emph{grayscale volume}). In essence, this step performs modality conversion and predicts how the same tissue will appear in visible light. 
	
	\subsubsection{Voxel aligned grayscale volume}
	In a real scenario a cryo volume for an MRI will not be available, therefore to generalize our method we train a neural network that learns to convert an MRI to a grayscale volume. In order to train such a network, both an MRI and a corresponding grayscale volume are needed. Unfortunately, a grayscale cryo volume is not suitable for this purpose due to the following reasons:
	\begin{itemize}
	    \item Even after careful registration, the features in cryo and MRI do not align perfectly, and there are errors of few voxels at some places, and
	    \item The MRI volume has a lower level of detail in comparison with the cryo volume that manifests as ambiguity in the tissue boundaries.
	\end{itemize}
	
	In order to create a grayscale volume that matches the features and scale of the MRI volume, we perform a Poisson reconstruction \cite{perez2003poisson}. Such a reconstruction fuses MRI gradients with gray intensities of the registered cryo volume.
	
	For visualization background removal is usually required. Therefore, we remove the background of the Poisson reconstruction using the segmentation mask. We then generate about 3000 sub-volumes of size $32\times32\times32$ for training the network. These volumes are taken from the MRI and the Poisson grayscale volumes by performing random rotations and translations in order to augment the training dataset.
	
	\subsubsection{GAN network for modality conversion}
	We use the sub-volumes generated from the MRI and Poisson grayscale volume to train a GAN intended to perform modality conversion of any MRI volume. The GAN learns to generate a  grayscale volume corresponding to an input MRI. The generator is based on 3D-FCN (Fully Convolutional Network)  without any max pooling layers. Since the MRI and grayscale volume have a strict geometric correspondence we avoid using max pooling layers. The generator architecture consists of 9 3D convolutional layers with filter size $3 \times3 \times 3$ and output feature maps of the following order 64, 128, 256, 512, 256, 128, 64, 1. Each subsequent convolution is followed with batch normalization and dropout layers. Leaky ReLU is used as activation except for the last layer where sigmoid is used.
	
\begin{figure}[htp]
	    \centering
	    \begin{tabularx}{\textwidth}{cXc}
	    \begin{tabular}{cccc}
	    & MRI & Grayscale & GT\\
	    \raisebox{0.25in}{\rotatebox{90}{\textbf{Axial}}} &
	    \includegraphics[angle=0,width=0.136\textwidth]{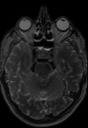} & 
	    \includegraphics[angle=0,width=0.136\textwidth]{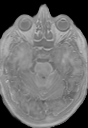} &
        \includegraphics[angle=0,width=0.136\textwidth]{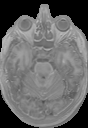}\\
        \raisebox{0.1in}{\rotatebox{90}{\textbf{Coronal}}}&
        \includegraphics[angle=0,width=0.136\textwidth]{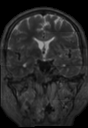} &
        \includegraphics[angle=0,width=0.136\textwidth]{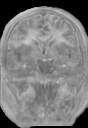} &
        \includegraphics[angle=0,width=0.136\textwidth]{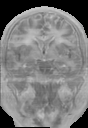}
	    \end{tabular}
	    & &
	    \begin{tabular}{ccc}
	       Style image & & \\
    	   \includegraphics[width=0.2\textwidth]{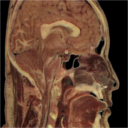} & \raisebox{0.25in}{\rotatebox{90}{Grayscale}} &  
           \includegraphics[width=0.2\textwidth]{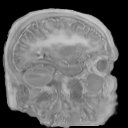}\\
           & \raisebox{0.25in}{\rotatebox{90}{Stylized}} &
           \includegraphics[width=0.2\textwidth]{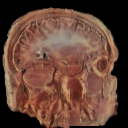}
	    \end{tabular}\\
	    (a) MRI to grayscale conversion. & & (b) Hint generation via stylization.\\
	    & & 
	\end{tabularx}
	\caption{Modality conversion and hint slice generation for colorization.}
	\label{fig:pipeline_stages_1_2}
	\end{figure}

	In order to ensure better structures we use Structural Similarity (SSIM) loss \cite{wang2004image} and an $L_1$ loss term. The SSIM loss enables the network to learn corresponding cryo structures more robustly while the $L_1$ loss helps in further improving the intensities of the generated voxels. The SSIM loss $\mathcal{L}_s$ between two volumes $I_{1}$ and $I_{2}$ is given by,
	\begin{align*}
	\mathcal{L}_s(I_{1},I_{2})  = 1 - \frac{(2\mu_1\mu_2 + c_1)(2\sigma_{1,2} + c_2)}{(\mu_1^2+\mu_2^2 + c_1)(\sigma_1^2 + \sigma_2^2 + c_2)},
	\end{align*}
	where $\mu_1$ and $\sigma_1^2$ are the mean and variance of $I_1$, $\mu_2$  and $\sigma_2^2$ are the mean and variance of $I_2$ and $\sigma_{1,2}$ is the covariance of $I_{1}$ and $I_{2}$. $c_1$ and $c_2$ are constants to prevent overflow in case of small denominators based on the dynamic range of voxel values. The complete generator loss can be formulated as,
	\begin{align}
	\mathcal{L}_G = \mathcal{L}_s(G(I_{mri}),I_{gray}) + ||G(I_{mri}) - I_{gray}||_1 + \log(1 - D(G(I_{mri}))),
	\label{eqn:gen_loss}
	\end{align}
	where $I_{mri}$ is the input MRI subvolume, $I_{gray}$ is the corresponding ground truth Poisson grayscale subvolume, $\log(1-D(G(I_{mri}))$ \cite{goodfellow2014generative} is the generator loss as described in  equation~(\ref{eqn:gen_loss}) for generator G and discriminator D for input volume. The discriminator loss can be formulated as, $\mathcal{L}_D = \log(D(G(I_{mri})))$. Our discriminator architecture is as following:
	\begin{center}
	\begin{tabularx}{0.9\textwidth}{cXcXcXc}\toprule
	     \textbf{Layer} & & \textbf{Filter size} & &\textbf{Stride} & &\textbf{kernel size}\\ \midrule
	     Conv3D + Batchnorm & & $1 \times 256$ & & 2 & & 3 \\
	     Conv3D + Batchnorm & & $256 \times 256$ & & 1 & & 3 \\
	     Conv3D + Batchnorm & & $256 \times 128$ & & 2 & & 3 \\
	     Conv3D + Batchnorm & & $128 \times 64$ & & 4 & & 3 \\ \bottomrule
	\end{tabularx}
	\end{center}
	The last layer being linear of size $512 \times 1$ with leaky relu as an activation function for all layers except last where sigmoid is used. The results of MRI to grayscale conversion are shown in Fig.~\ref{fig:pipeline_stages_1_2}(a). The performance of our network can be seen by its ability to generate a grayscale volume that is as close as possible to the ground truth (GT) Poisson grayscale volume.
	
	\subsection{Style transfer for generating hints}
	For generating planar hints, we use the style transfer method proposed by Gatys et al. \cite{gatys2016image}. The objective is to minimize the style loss and content loss, for which convolution layers of a pre-trained VGG19\cite{vgg} are used. We use the features from the first five convolutions to characterize the style and content loss. Fig.~\ref{fig:pipeline_stages_1_2}(b) shows sample colorization.
	
    When we apply Gatys style transfer on the complete volume slice by slice, it leads to incoherent results because the loss function is defined for images and it ignores the content and style information along the third dimension in the volume. Therefore, as explained next we utilize a separate colorization step that completely preserves the MRI structures while propagating colors from the hints.

	\subsection{Volume colorization using hints}
	To perform colorization we refer to the work by Levin et al. \cite{levin2004colorization} that performs colorization using optimization with hints given by the user. We extend this method to 3D volume colorization and use the stylized images as 2D hints for colorization in 3D. To select slices as hints, we choose the slices along any of the axes with a high standard deviation to ensure good contrast. The colorization is solved as a constrained optimization problem where two neighboring voxels in a volume, $p$ and $q$ should have the same colors if their intensities are similar. The colorization is performed in the $YUV$ color space where the $Y$ channel is the same as the one generated using GAN and the cost function for $U$ and $V$ is defined as
	\begin{align*}
	J(U) = \sum_p \bigg( U(p) - \sum_{q \in N(p)} w_{pq}U(q) \bigg)^2
	\end{align*}
	where $N(p)$ is a set of all voxels that lie in a $3 \times 3 \times 3$ window around $p$. $w_{pq}$ is a weighting function which is large for similar values of $Y(p)$ and $Y(q)$. Here we used $w_{pq} = \displaystyle e^{- (Y(p) - Y(q))^2 / 2 \sigma^2_p}$, where $\sigma_p$ is the the variance along $N(p)$. Given the quadratic cost function subject to the defined constraints, we minimize $J(U)$ and $J(V)$ . Since the optimization problem creates a large system of sparse equations, we use multi-grid solver \cite{OlSc2018} to obtain an optimal solution.
	
    \begin{figure}[htp]	
	    \centering
	    \begin{tabular}{c c c c c}
	         {Style image} & & \textbf{Axial} & \textbf{Coronal} & \textbf{Sagittal}\\
	         \includegraphics[width=0.2\textwidth]{./gatys_style_image_2_resize.png} &
	         \raisebox{0.2in}{\rotatebox{90}{Input MRI }}  &
	         \includegraphics[width=0.136\textwidth]{./mrgr_l_axial.png}&  
	         \includegraphics[width=0.136\textwidth]{./mrgr_l_coronal.png}& 
	         \includegraphics[width=0.2\textwidth]{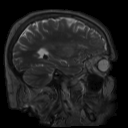}\\
             & \raisebox{.1in}{\rotatebox{90}{Colorization}} &
	         \includegraphics[width=0.136\textwidth]{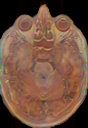} & 
	         \includegraphics[width=0.136\textwidth]{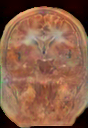} &
	         \includegraphics[width=0.2\textwidth]{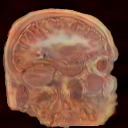}
	    \end{tabular}
	  
	    \caption{Colorization result with a cryo style image.}
	    \label{fig:sample_opt_colorized}
	\end{figure} 

	\section{Results and comparisons}
	We train our method using the Visible Human 2 \cite{ackerman1998visible} 16-bit MRI and color cryo dataset. Our code is written in Python where we use PyTorch \cite{NEURIPS2019_9015} for training the GAN and PyAMG \cite{OlSc2018} for solving the optimization problem to generate the colorized volume. The pipeline is run on an Intel Xeon 2.4 GHz processor with 128 GB memory and Nvidia Quadro P6000 GPU with 24 GB memory. Our code will be made available for research purposes.
		
\begin{figure}[htp]
	    \centering
	    \begin{tabular}{ccccccc}
        Style image & MRI & (a) & (b) & (c) & (d) & (e) \\
	    \includegraphics[angle=0,width=0.136\linewidth]{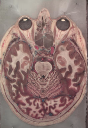} &
	    \includegraphics[angle=0,width=0.136\linewidth]{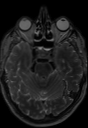} &
        \includegraphics[angle=0,width=0.136\linewidth]{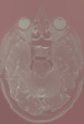} &
        \includegraphics[angle=0,width=0.136\linewidth]{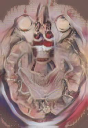} &
        \includegraphics[angle=0,width=0.136\linewidth]{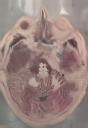} &
        \includegraphics[angle=0,width=0.136\linewidth]{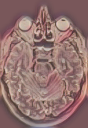} &
        \includegraphics[angle=0,width=0.136\linewidth]{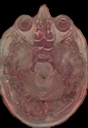} \\
        &
        \includegraphics[angle=0,width=0.136\linewidth]{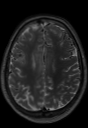} &
        \includegraphics[angle=0,width=0.136\linewidth]{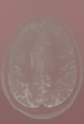} &
        \includegraphics[angle=0,width=0.136\linewidth]{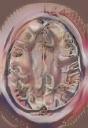} &
        \includegraphics[angle=0,width=0.136\linewidth]{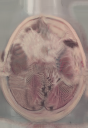} &
        \includegraphics[angle=0,width=0.136\linewidth]{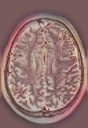} &
        \includegraphics[angle=0,width=0.136\linewidth]{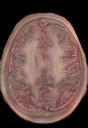}
	    \end{tabular}
	    \caption{Comparison with recent style transfer methods: (a) Li et al. \cite{li2018closed}, (b) Huang et al. \cite{huang2017arbitrary}, (c) Jamrivska et al. \cite{jamrivska2019stylizing}, (d) Li et al. \cite{li2019learning}, (e) Ours.}
	    \label{fig:comparison}
	\end{figure}

    We show results of our colorization method on MRI volumes in Fig.~\ref{fig:sample_opt_colorized} and Fig.~\ref{fig:comparison}(e). It can be seen that the tissue intensities in MRI have been correctly mapped to tissue colors of the cryo style image, and structures of input MRI are maintained very well. In general MRI intensities do not linearly map to cryo intensities and our colorization takes care of such differences in the two modalities.
	
    To the best of our knowledge, there are no existing methods that can be compared directly with our method. Most existing colorization methods work on 2D photographs and none on medical images. Some modality conversion approaches are also not directly comparable to our color synthesis. We compare our approach with four state-of-the-art methods after suitable modifications needed to be applied for medical imaging data (see Fig.~\ref{fig:comparison}). All of these methods require 2D images as input and therefore we process the MRI volume slice-by-slice along the axial plane. In practice, we suggest a full 3D processing as with our algorithm since that leads to a better coherency in the volume compared to a slice-by-slice approach.
   
   In the comparisons, it can be seen that Huang et al. \cite{huang2017arbitrary} is unable to maintain coherency across different slices and also adds undesirable artifacts. Li et al. \cite{li2018closed} also do not maintain such a coherency. In addition, they perform photorealistic smoothening on the stylized image that blurs the structures of the input volume. Since Li et al. \cite{li2019learning} operates on a video sequence, it preserves coherency across slices but does not fully transfer the style details. Results of Jamrivska et al. \cite{jamrivska2019stylizing} are best at reproducing the style but the method requires hints that are completely aligned with the input slices that may not be available for every style image. It also fails to preserve the structures in the input frames. We also perform a quantitative comparison of the generated images with respect to the input MRI. Table~\ref{tab:tbl_metrics} shows average SSIM, PSNR, and MSE metrics computed for the entire volume and averaged over RGB color channels.

	\begin{table}[htp]
		\caption{Quantitative comparison with recent style transfer methods}
	    \label{tab:tbl_metrics}
	    \centering
	    \begin{tabular}{p{4cm}ccc}\toprule
	        \textbf{Method} & \textbf{SSIM $\uparrow$} & \textbf{PSNR $\uparrow$} & \textbf{MSE $\downarrow$} \\ \midrule
	        Li et al. \cite{li2018closed} & 0.313 & 20.18 & 0.165 \\
	        Huang et al. \cite{huang2017arbitrary} & 0.313 & 20.15 & 0.194 \\
	        Jamrivska et al. \cite{jamrivska2019stylizing} & 0.312 & 20.10 & 0.257 \\
	        Li et al. \cite{li2019learning}  & 0.314 & 20.19 & 0.153 \\
	        Ours & \textbf{0.319} & \textbf{21.81} & \textbf{0.052} \\ \bottomrule
	\end{tabular}
	\end{table}
	
	\section{Discussion and Challenges}
	The unique setting of our problem brings forward several key challenges in medical data colorization. The 3D nature of the data places severe restrictions on computational resources. The difference in modality leads to a lack of direct translation between MRI and cryo volume, thus requiring the need for grayscale synthesis. A key limitation of our method is that the lack of semantic information for MRI volume prevents appropriate stylization of hints. Gatys style transfer method and other stylization techniques are not suitable for medical data and do not seem to perform well leading to issues such as patchy colors and color bleeding due to lack of semantic information. Further, the lack of suitable colorization metrics for medical images make comparison and evaluation difficult. 
	
	\section{Conclusion}
	Overall we find that our approach is able to handle the wide differences present in the MRI and the color modalities. Our experiments also indicate that GANs provide a well-suited framework for the colorization of such medical data. This research opens up many broad aspects of multi-modal colorization that has applications for high-fidelity photorealistic visualization and colorization.
	
	%
	%
	\bibliographystyle{splncs04}
	\bibliography{arxiv20colorization}

\begin{thebibliography}{10}
\providecommand{\url}[1]{\texttt{#1}}
\providecommand{\urlprefix}{URL }
\providecommand{\doi}[1]{https://doi.org/#1}

\bibitem{ackerman1998visible}
Ackerman, M.J.: The visible human project. Proceedings of the IEEE
  \textbf{86}(3),  504--511 (1998)

\bibitem{gatys2016image}
Gatys, L.A., Ecker, A.S., Bethge, M.: Image style transfer using convolutional
  neural networks. In: Proceedings of the IEEE Conference on Computer Vision
  and Pattern Recognition. pp. 2414--2423 (2016)

\bibitem{goodfellow2014generative}
Goodfellow, I., Pouget-Abadie, J., Mirza, M., Xu, B., Warde-Farley, D., Ozair,
  S., Courville, A., Bengio, Y.: Generative adversarial nets. In: Advances in
  neural information processing systems. pp. 2672--2680 (2014)

\bibitem{guibas2017synthetic}
Guibas, J.T., Virdi, T.S., Li, P.S.: Synthetic medical images from dual
  generative adversarial networks. arXiv preprint arXiv:1709.01872  (2017)

\bibitem{hertzmann2001image}
Hertzmann, A., Jacobs, C.E., Oliver, N., Curless, B., Salesin, D.H.: Image
  analogies. In: Proceedings of the 28th annual conference on Computer graphics
  and interactive techniques. pp. 327--340. ACM (2001)

\bibitem{huang2017arbitrary}
Huang, X., Belongie, S.: Arbitrary style transfer in real-time with adaptive
  instance normalization. In: Proceedings of the IEEE International Conference
  on Computer Vision. pp. 1501--1510 (2017)

\bibitem{iizuka2016let}
Iizuka, S., Simo-Serra, E., Ishikawa, H.: Let there be color!: joint end-to-end
  learning of global and local image priors for automatic image colorization
  with simultaneous classification. ACM Transactions on Graphics (TOG)
  \textbf{35}(4), ~110 (2016)

\bibitem{jamrivska2019stylizing}
Jamri{\v{s}}ka, O., Sochorov{\'a}, {\v{S}}., Texler, O., Luk{\'a}{\v{c}}, M.,
  Fi{\v{s}}er, J., Lu, J., Shechtman, E., S{\`y}kora, D.: Stylizing video by
  example. ACM Transactions on Graphics (TOG)  \textbf{38}(4),  1--11 (2019)

\bibitem{levin2004colorization}
Levin, A., Lischinski, D., Weiss, Y.: Colorization using optimization. In: ACM
  transactions on graphics (tog). vol.~23, pp. 689--694. ACM (2004)

\bibitem{li2019learning}
Li, X., Liu, S., Kautz, J., Yang, M.H.: Learning linear transformations for
  fast image and video style transfer. In: Proceedings of the IEEE Conference
  on Computer Vision and Pattern Recognition. pp. 3809--3817 (2019)

\bibitem{li2017universal}
Li, Y., Fang, C., Yang, J., Wang, Z., Lu, X., Yang, M.H.: Universal style
  transfer via feature transforms. In: Advances in neural information
  processing systems. pp. 386--396 (2017)

\bibitem{li2018closed}
Li, Y., Liu, M.Y., Li, X., Yang, M.H., Kautz, J.: A closed-form solution to
  photorealistic image stylization. In: Proceedings of the European Conference
  on Computer Vision (ECCV). pp. 453--468 (2018)

\bibitem{ljung2016state}
Ljung, P., Kr{\"u}ger, J., Groller, E., Hadwiger, M., Hansen, C.D., Ynnerman,
  A.: State of the art in transfer functions for direct volume rendering. In:
  Computer Graphics Forum. vol.~35, pp. 669--691. Wiley Online Library (2016)

\bibitem{nie2017medical}
Nie, D., Trullo, R., Lian, J., Petitjean, C., Ruan, S., Wang, Q., Shen, D.:
  Medical image synthesis with context-aware generative adversarial networks.
  In: International Conference on Medical Image Computing and Computer-Assisted
  Intervention. pp. 417--425. Springer (2017)

\bibitem{nie2018medical}
Nie, D., Trullo, R., Lian, J., Wang, L., Petitjean, C., Ruan, S., Wang, Q.,
  Shen, D.: Medical image synthesis with deep convolutional adversarial
  networks. IEEE Transactions on Biomedical Engineering  \textbf{65}(12),
  2720--2730 (2018)

\bibitem{nie2007optimization}
Nie, D., Ma, Q., Ma, L., Xiao, S.: Optimization based grayscale image
  colorization. Pattern recognition letters  \textbf{28}(12),  1445--1451
  (2007)

\bibitem{OlSc2018}
Olson, L.N., Schroder, J.B.: {PyAMG}: Algebraic multigrid solvers in {Python}
  v4.0 (2018), \url{https://github.com/pyamg/pyamg}, release 4.0

\bibitem{NEURIPS2019_9015}
Paszke, A., Gross, S., Massa, F., Lerer, A., Bradbury, J., Chanan, G., Killeen,
  T., Lin, Z., Gimelshein, N., Antiga, L., Desmaison, A., Kopf, A., Yang, E.,
  DeVito, Z., Raison, M., Tejani, A., Chilamkurthy, S., Steiner, B., Fang, L.,
  Bai, J., Chintala, S.: Pytorch: An imperative style, high-performance deep
  learning library. In: Wallach, H., Larochelle, H., Beygelzimer, A.,
  d\textquotesingle Alch\'{e}-Buc, F., Fox, E., Garnett, R. (eds.) Advances in
  Neural Information Processing Systems 32, pp. 8024--8035. Curran Associates,
  Inc. (2019),
  \url{http://papers.neurips.cc/paper/9015-pytorch-an-imperative-style-high-performance-deep-learning-library.pdf}

\bibitem{perez2003poisson}
P{\'e}rez, P., Gangnet, M., Blake, A.: Poisson image editing. ACM Transactions
  on graphics (TOG)  \textbf{22}(3),  313--318 (2003)

\bibitem{vgg}
Simonyan, K., Zisserman, A.: Very deep convolutional networks for large-scale
  image recognition. arXiv preprint arXiv:1409.1556  (2014)

\bibitem{tu2009auto}
Tu, Z., Bai, X.: Auto-context and its application to high-level vision tasks
  and 3d brain image segmentation. IEEE transactions on pattern analysis and
  machine intelligence  \textbf{32}(10),  1744--1757 (2009)

\bibitem{wang2004image}
Wang, Z., Bovik, A.C., Sheikh, H.R., Simoncelli, E.P., et~al.: Image quality
  assessment: from error visibility to structural similarity. IEEE transactions
  on image processing  \textbf{13}(4),  600--612 (2004)

\bibitem{welsh2002transferring}
Welsh, T., Ashikhmin, M., Mueller, K.: Transferring color to greyscale images.
  In: ACM Transactions on Graphics (TOG). vol.~21, pp. 277--280. ACM (2002)

\bibitem{zhang2016colorful}
Zhang, R., Isola, P., Efros, A.A.: Colorful image colorization. In: European
  conference on computer vision. pp. 649--666. Springer (2016)

\bibitem{zhang2017real}
Zhang, R., Zhu, J.Y., Isola, P., Geng, X., Lin, A.S., Yu, T., Efros, A.A.:
  Real-time user-guided image colorization with learned deep priors. arXiv
  preprint arXiv:1705.02999  (2017)

\end{thebibliography}
\end{document}